\documentclass[a4paper]{article}

\usepackage{INTERSPEECH2021}
\usepackage{cite}

\title{Speech Activity Detection Based on Multilingual Speech Recognition System}
 \name{Seyyed Saeed Sarfjoo, Srikanth Madikeri, and Petr Motlicek}
\address{Idiap Research Institute, Martigny, Switzerland}
\email{\{ssarfjoo, msrikanth, petr.motlicek\}@idiap.ch}

\begin{document}

\maketitle
\begin{abstract}
To better model the contextual information and increase the generalization ability of Speech Activity Detection (SAD) system, this paper leverages a multi-lingual Automatic Speech Recognition (ASR) system to perform SAD. Sequence-discriminative training
of Acoustic Model (AM) using Lattice-Free Maximum Mutual Information (LF-MMI) loss function, effectively extracts the
contextual information of the input acoustic frame. Multi-lingual AM training, causes the robustness to noise and language variabilities. The index of maximum output posterior is considered as a frame-level speech/non-speech decision function. Majority voting and logistic regression are applied to fuse the language-dependent decisions. The multi-lingual ASR is trained on 18 languages of BABEL datasets and the built SAD is evaluated on 3 different languages. On out-of-domain datasets, the proposed SAD model shows significantly better performance with respect to baseline models. On the Ester2 dataset, without using any in-domain data, this model outperforms the WebRTC, phoneme recognizer based VAD (Phn\_Rec), and Pyannote baselines (respectively by 7.1, 1.7, and 2.7\% absolute) in Detection Error Rate (DetER)
metrics. Similarly, on the LiveATC dataset, this model outperforms the WebRTC, Phn\_Rec, and Pyannote baselines (respectively by 6.4, 10.0, and 3.7\% absolutely) in DetER metrics.
\end{abstract}


\section{Introduction}
\label{seq:intro}
Speech Activity Detection (SAD), a process of identifying the speech segments in an audio utterance~\cite{sohn1999statistical}, is a critical part of Automatic Speech Recognition (ASR), speaker recognition, speaker diarization, and other speech-based applications. Developing an accurate SAD system, operating in the noisy environment is an active research field in speech processing~\cite{sharma2019multi,martinelli2020spiking,dellaferrera2020bin,lee2020dual,zheng2020mlnet}.

This paper explores SAD built around multi-lingual ASR systems, as we hypothesize it can offer better \emph{generalization ability} by leveraging the contextual information extracted by ASR~\cite{madikeri2020lattice}.
%In this paper, for better extracting the contextual information and \emph{generalization ability} of the trained SAD model, the SAD model was built over the multi-lingual ASR system~\cite{madikeri2020lattice}. Multi-lingual acoustic model training combines data from multiple languages to train the ASR model.
Generally, this paper employs a conventional multi-task network as a multi-lingual Acoustic Model (AM) trained using the Lattice-Free Maximum Mutual Information (LF-MMI) framework, capable to  %  using a scale-able approach
%to train the multi-lingual acoustic model is capable to 
extract the language-dependent contextual information. 
Using a multi-lingual dataset for the AM training was investigated in~\cite{karafiat2020but}. Unlike applying a simple block-softmax loss on stacked input data with added language indicator for
phoneme names, we apply LF-MMI loss on multi-task architecture which provides a scalable approach to develop multi-lingual AM. Practically, we use PKWRAP, a PyTorch package for LF-MMI training of acoustic models~\cite{madikeri2020pkwrap}. The proposed multi-lingual acoustic model was trained on 18 languages of the BABEL datasets. The original motivation for using this dataset is to train an SAD system robust to noise and language variabilities. Within each language-dependent part of AM, speech and non-speech acoustic frames are mapped to a different set of output context-dependent phones (i.e. posteriors). For each language, we use index of maximum output posterior as a frame-level speech/non-speech decision function. To fuse the decisions from different languages, conventional logistic regression~\cite{kleinbaum2002logistic} and majority voting techniques are employed. 

To investigate the generalization ability of the proposed SAD, experiments presented in the paper were performed on both in-domain and out-of-domain data. For out-of-domain experiments, two specific conditions are considered: (i) an access to small development set is available, or (ii) no in-domain data is available at all. Logistic regression and majority voting fusion are reported for these conditions. 
Concretely, the development part of the BABEL Kurdish dataset is used as an in-domain evaluation set. Eval parts of Ester2\footnote{http://catalog.elra.info/en-us/repository/browse/ELRA-S0338/} and  LiveATC\footnote{https://www.liveatc.net/} datasets are used as out-of-domain sets. BABEL Kurdish contains conversational telephony speech (CTS) in the Kurdish. Ester2 is a broadcast news dataset in French. LiveATC comprises large number of conversations between Air Traffic Controllers (ATCo) and pilots with a large variety of accents in English. To investigate the generalization ability of our SAD model, we consider different real-life scenarios with high variation in channel, background noise, and language. Having a generic and robust SAD is critical for downstream tasks such as ASR.

We show that the proposed multi-lingual architecture offers comparable results on in-domain set and significantly outperforms the baselines on out-of-domain Ester2 and LiveATC datasets. For fair comparison with Google WebRTC and popular BUT pre-trained phoneme recognizer based SAD (Phn\_Rec)\footnote{https://speech.fit.vutbr.cz/software/phoneme-recognizer-based-long-temporal-context} in out-of-domain evaluation, we also assumed that no in-domain data is available during training. In addition, using a small development set in the logistic regression method further improves the performance of the proposed SAD system.

The rest of this paper is organized as follows: Related works are shown in Section~\ref{sec:relatedwork}. Multi-lingual acoustic model training is briefly explained in Section~\ref{sec:mulasr}. The proposed multi-lingual ASR-based SAD is described in Section~\ref{sec:mulasrSAD}. Experiment setup and results are shown in Section~\ref{sec:experiments}. Conclusions are discussed in Section~\ref{sec:conclusions}.

\section{Related works}
\label{sec:relatedwork}
Large effort was invested in the past to find the optimal features~\cite{chuangsuwanich2011robust, misra2012speech, zazo2016feature}, or classifier~\cite{ng2012developing,veisi2012hidden,enqing2002applying} for the SAD task. We can mention Gaussian Mixture Model (GMM)~\cite{ng2012developing}, Hidden Markov Model (HMM)~\cite{veisi2012hidden}, or Support Vector Machines (SVM)~\cite{enqing2002applying} as the often used classifiers for the SAD task. With the advent of Deep Neural Networks (DNNs), several DNN-based architectures were proposed for the SAD task~\cite{zhang2012deep,ivry2019voice} including Convolutional Neural Network (CNN)~\cite{chang2018temporal} and Recurrent Neural Network (RNN)~\cite{hughes2013recurrent} architectures. Recently, for training the SAD model in a noisy environment, DNN models with attention mechanism in temporal domain~\cite{zheng2020mlnet} and a combination of temporal and spectral domains~\cite{lee2020dual} were investigated.

Contextual Information (CI) is important for training a robust SAD system, specially at low Signal-to-Noise Ratios (SNR)~\cite{zhang2015boosting}. Several methods for boosting the contextual information have been proposed. In~\cite{zhang2014boosted}, by boosting CI, Zhang and Wang proposed to generate multiple different predictions from a single DNN and reported significant improvement over the standard DNN in challenging noise scenarios with low SNR levels. In~\cite{zhang2015boosting}, a boosted DNN (bDNN)-based SAD was proposed. Zhang and Wang exploited the input/output CI by adopting multiple input/output units for the DNN. In addition, to aggregate long-short term CI, they proposed an ensemble model that contains bDNNs of various sizes. However, the computational cost of ensembled method is significantly higher than of a single-bDNN-based SAD. 

Capturing sequential contextual information using RNN architecture was investigated in~\cite{hughes2013recurrent}, nevertheless, the improvement in the results were observed when the models were trained as \emph{noise-dependent}. Using multi-lingual BABEL or Public Safety Communications (PSC) datasets for training the DNN based SAD with simple feed-forward architecture was investigated in~\cite{karafiat2020but}. PSC corpus that contains simulated first-responder type background noises and speech affects, was introduced in NIST OpenSAT 2019 challenge~\cite{byers2019open}. Similar to LiveATC, this dataset is challenging for ASR and SAD tasks.

% \vspace{-5mm}
\section{Multi-lingual acoustic model training }
\label{sec:mulasr}
Training multi-lingual ASR system is an effective way to compensate data shortage in low-resourced languages. DNN based acoustic models can be considered as a feature extractor to train a monolingual acoustic model for the specific target language. The multi-lingual models can either share the output layer or have separate output
layers, which are called single- and multi-task models, respectively. Without any loss in performance, multi-task ASR training provides a much more scalable approach to develop multi-lingual AM~\cite{madikeri2020lattice}. LF-MMI significantly outperformed the conventional cross-entropy (CE) for training the multi-lingual AM~\cite{hadian2018end}. The MMI cost function uses a numerator and a denominator graph to model the observed feature sequence based on the ground truth and compute the probability over all possible sequences, respectively. Sequence-discriminative training
of multi-lingual AM using LF-MMI loss function, effectively extracts the contextual information of the input acoustic frame. In this paper, for training the multi-lingual AM, time delayed neural network (TDNN) architecture with LF-MMI loss was applied. In order to obtain alignments to train all the TDNN models, HMM/GMM models were first trained for each language. 

In multi-task training of AM, we have $L$ objective functions where $L$ is the number of training languages, computed independent of each other based on the language of the input utterance:

\begin{equation}
\mathcal{F}_{\text{MMI}}^{(l)} = \sum_{u=1}^{U_{l}} \log \frac{p\big(\textbf{x}^{(u)}|\mathcal{M}_{\textbf{w}(u)}^{l},\theta \big)  p(\textbf{w}(u))}{p\big(\textbf{x}^{(u)}|\mathcal{M}_{\text{den}}^{l},\theta\big)},
\label{eq:mullfmmi}
\end{equation}

where $U_{l}$ is the number of utterances in the current minibatch for language $l$, $\bf{\theta}$ contains the shared and language-dependent parameters, $\mathcal{M}_{\textbf{w}(u)}^{l}$ and $\mathcal{M}_{\text{den}}^{l}$ are language-specific numerator and denominator graphs, respectively. The overall cost function is the weighted sum of all language-dependent cost functions:

\begin{equation}
\mathcal{F}_{\text{MMI}} = \sum_{l=1}^{L} \alpha_{l}\mathcal{F}_{\text{MMI}}^{(l)},
\label{eq:mullancost}
\end{equation}
 
where $\alpha_{l}$ is the language-dependent weight for computing the total loss. Gradients for language-dependent layers are computed and updated for each minibatch. Using backpropogation, the shared parameters are then updated. 

% \vspace{-3mm}
\section{Multi-lingual ASR based SAD}
\label{sec:mulasrSAD}

\begin{figure}[h]

    \centering
    \includegraphics[width=0.45\textwidth, trim={2cm, 1cm, 7cm, 1.1cm}, ]{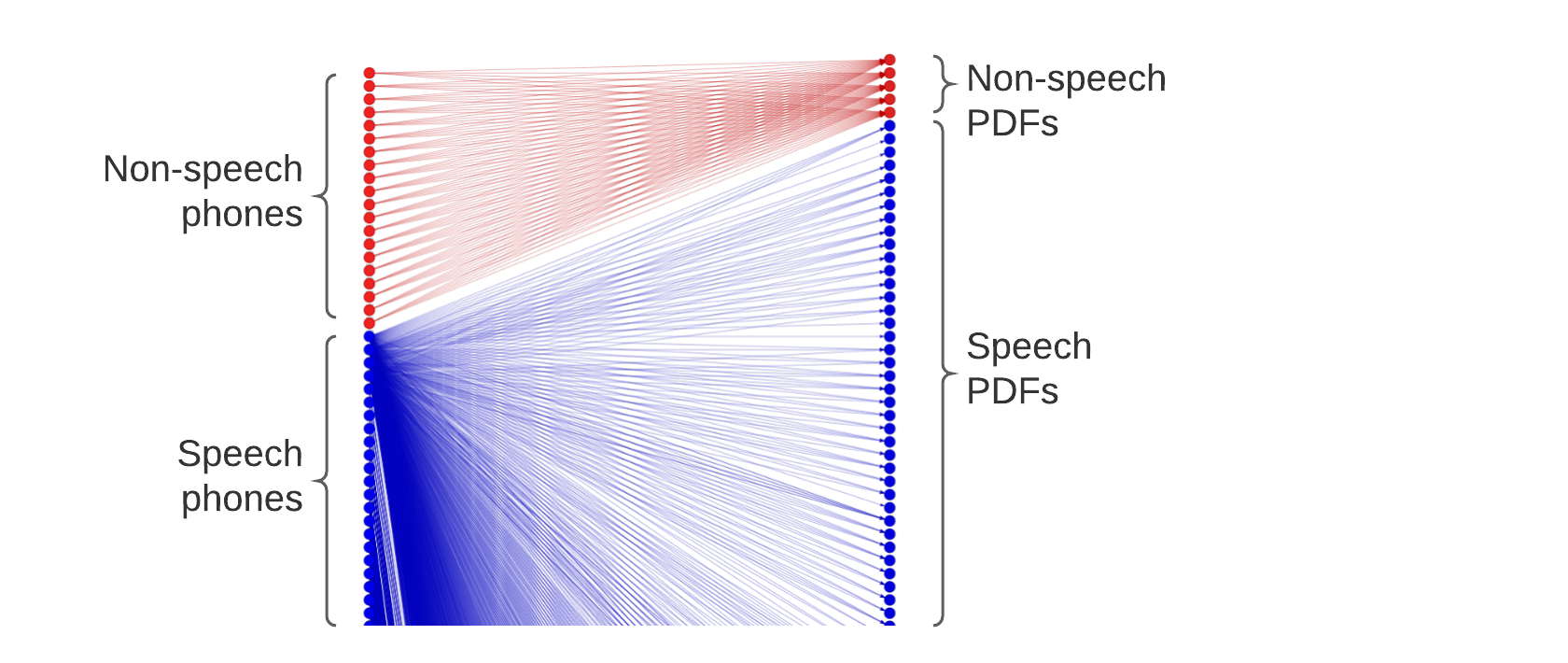}
    \caption{Mapping between input phones and output PDFs in the HMM/GMM ASR model of Assamese language. Non-speech phones are mapped to the first five initial PDFs.}
    \label{fig:mapphone2pdf}
\end{figure}

In the trained HMM/GMM model for each language, we can observe the mapping between input phones and the output Probability Density Functions (PDFs). Figure~\ref{fig:mapphone2pdf}, shows the mapping between input phones and output PDFs in the HMM/GMM ASR model of BABEL Assamese language. We can observe that the non-speech phones are mapped to the specific non-speech PDFs. As a result, the output posteriors of the language-dependent AM model are separated for input speech and non-speech frames. For training the AM model, the sequential discriminative LF-MMI loss function was applied. As a result, these output posteriors can be effective for discriminating the speech and non-speech frames.

\begin{figure}[h]
    \centering
    \includegraphics[width=0.45\textwidth, trim={0cm, 1.8cm, 0cm, 0cm}, ]{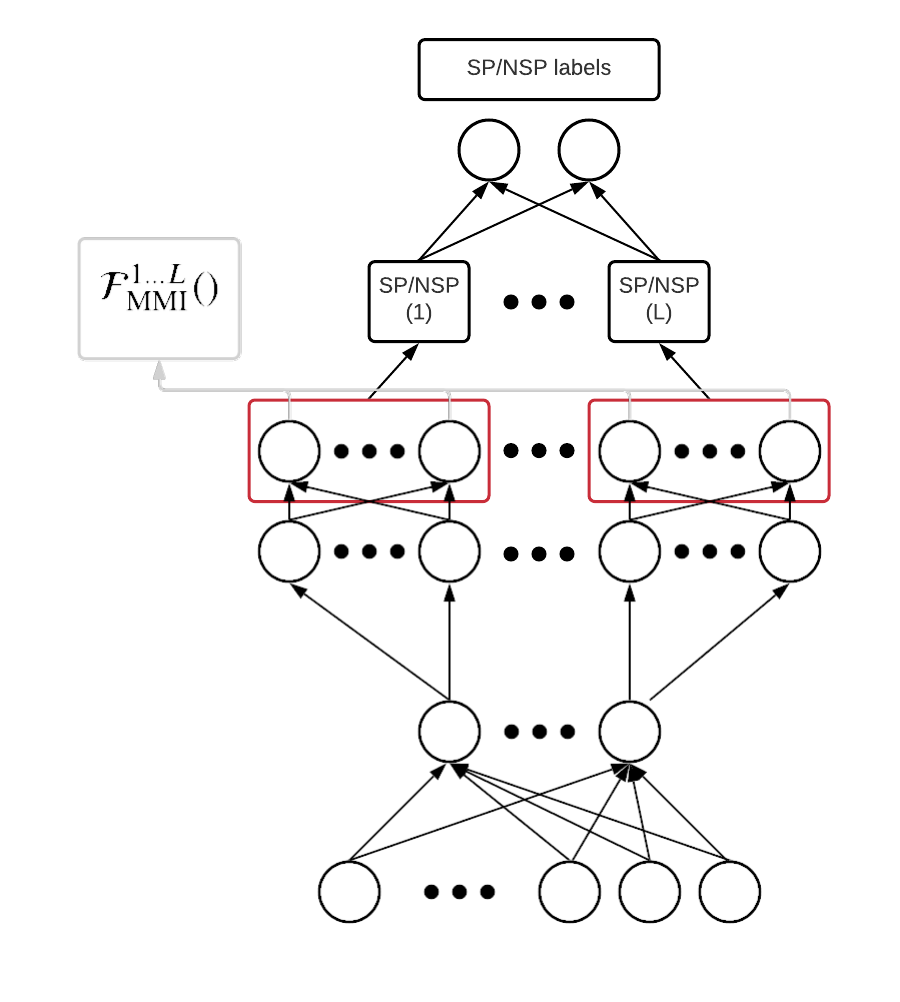}
    \caption{Structure of the multi-lingual ASR based SAD. SP/NSP block detects speech frames based on the PDF index of frame-level maximum output posteriors. Logistic regression was used for fusing the language dependent speech/non-speech labels.}
    \label{fig:mulasrSAD}
    
\end{figure}

The structure of multi-lingual ASR-based SAD is shown in the Figure~\ref{fig:mulasrSAD}. After training the multi-lingual AM model, for each language we consider the speech/non-speech (SP/NSP) block, to detect the speech frames based on the PDF index of frame-level maximum output posterior. If the maximum output posterior belongs to one of the non-speech PDFs, we consider the current frame as non-speech frame and vice versa. For fusing the decisions from different languages, we performed logistic regression and majority voting techniques. In logistic regression, we concatenate the frame-level decisions of SP/NSP block of each language as input features to predict the final decision. In the majority voting, we consider a frame as speech if SP/NSP block of majority of the languages consider it as speech frame.     

% \vspace{-3mm}
\section{Experimental setup and results}
\label{sec:experiments}

\subsection{Dataset and DNN configuration}
\label{subsec:dataset}

To demonstrate the scalability of the multi-task system, we consider training of the multi-lingual acoustic model with 18 languages from BABEL datasets with approximately 1000 hours of data. All languages are available at Linguistic Data Consorcium\footnote{https://www.ldc.upenn.edu} (LDC). The name of BABEL languages used for training is shown in the Table~\ref{tab:languagenames}.

\begin{table}[h]
\caption{BABEL languages used for training.}
\centering
% \resizebox{0.48\textwidth}{!}{%
\begin{tabular}{@{}c@{}}
\toprule
Languages \\
\midrule
Assamese, Bengali, Cantonese, Haitian, Kazhak, Kurmanji\_kurdish, \\
Lao, Lithuanian, Pashto, Somali, Swahili, Tagalog, \\
Tamil, Telugu, Tok\_pisin, Turkish, Vietnamese, Zulu \\
\bottomrule
\end{tabular}
%  }
\label{tab:languagenames}
\end{table}

For training the AM, we used 40-dimensional MFCCs as acoustic features, derived from 25 ms frames with a 10 ms frame shift. In addition, an online i-vector extractor of 100 dimension is trained. For speeding up the training, we used a frame sub-sampling factor of 3. We also augmented the data with 2-fold speed perturbation in all the experiments. The network consists of 8 layers of TDNN with 1024 nodes in each layer. The pre-final layer has only 200 units. For training the AM model, PKWRAP, a PyTorch package for LF-MMI training of acoustic models was used~\cite{madikeri2020pkwrap}. Real-time factor for extracting the MFCC, i-vector, and computing the DNN posteriors on 4.20GHz Intel(R) Core(TM) i7-7700K CPU are 0.0076, 0.0032, and 0.18, respectively. Real-time factor for computing the DNN posteriors on GeForce GTX 1080 Ti GPU is 0.0066. The time for computing the logistic regression scores w.r.t. forward pass of the DNN is negligible. Using CPU and GPU for processing, this model is roughly five and sixty two times faster than real-time. Because of this reason this approach is convenient as a pre-processing step for audio-based interactive systems.

For investigating the generalization ability of the proposed SAD, we performed experiments on in-domain and out-of-domain scenarios. Development part of BABEL Kurdish dataset was used as in-domain evaluation set. Eval parts of Ester2 and LiveATC datasets were used as out-of-domain sets. LiveATC was collected in the automatic collection and processing of voice data from air-traffic communications (ATCO2) project.\footnote{https://www.atco2.org/} For all evaluations, we consider the conditions when we have access to the in-domain development set, which is used for training the logistic regression (ASR\_Mul\_LR) and the condition which we do not have development set, which is the case for majority voting (ASR\_Mul\_MV) and SP/NSP blocks of single best language. For ASR\_Mul\_LR models, the threshold for SP/NSP detection was set based on Half Total Error Rate (HTER). The duration and number of segments in the selected datasets are shown in the Table\ref{tab:dataset}. For investigating the \emph{generalization ability} of our SAD model, we considered different real life scenarios with high variation in channel, background noise, and language.

\begin{table}[h]
\caption{Duration and number of segments in the selected datasets.}
\centering
% \resizebox{0.48\textwidth}{!}{%
\begin{tabular}{@{}lcc@{}}
\toprule
Dataset & Duration (hour) & \# Segments \\
\midrule
LiveATC\_dev & 2.7 & 1.0k \\ 
LiveATC\_eval & 6.8 & 0.9k \\ 
Ester2\_dev & 7.4 & 1.2k \\ 
Ester2\_eval & 7.2 & 1.7k \\ 
BabelKurdish\_dev & 20.6 & 11.0k \\ 
BabelKurdish\_eval & 20.0 & 11.3k \\ 
\bottomrule
\end{tabular}
%  }
\label{tab:dataset}
\end{table}

In this experiment, False Alarm (FA), Miss detection (Miss), and Detection Error Rate (DetER) were used as performance measures. DetER is defined as:

\begin{equation}
\text{DetER} = \frac{\text{False alarm} + \text{Miss detection}}{\text{Total duration of speech in the reference file}}.
\label{eq:deter}
\end{equation}

FA and Miss, are performance measures with just considering the $\text{False alarm}$ and $\text{Miss detection}$ in the numerator of $\text{DetER}$, respectively. In this paper, we considered Brno University's phoneme recognizer based VAD (Phn\_Rec), Google webRTC\footnote{https://github.com/wiseman/py-webrtcSAD}, and Bi-directional LSTM (BLSTM) based SAD from Pyannote~\cite{Bredin2020} as baseline models. Phn\_Rec is Hungarian phoneme recognizer, with all the phoneme classes linked to 'speech' class~\cite{schwarz2006hierarchical}. Hungarian data which was collected in SpeechDat-E project\footnote{http://www.fee.vutbr.cz/SPEECHDAT-E}, was found as the best for generic phoneme recognition working over the different languages~\cite{matejka2007but}. For Pyannote SAD, we trained the model using the same 18 BABEL languages. For fair comparison with WebRTC and Phn\_Rec, we considered the out-of-domain scenarios when we don't have access to the in-domain data. Based on the SAD result on the second DIHARD challenge\footnote{https://dihardchallenge.github.io/dihard2}, the aggressiveness mode of WebRTC SAD was set to 3.  

\vspace{-1mm}
\subsection{In-domain evaluation}
\label{subsec:indomain}

Comparison of SAD results on in-domain experiment for BabelKurdish\_eval set is shown in the Table~\ref{tab:SADresultskurdish}. To reduce the noise in the classifier’s output, in each ASR based SAD, we applied the temporal smoothing for detecting the start and end of each speech segment. In all experiments output of Tok\_pisin language showed single best result which is ASR\_Single\_Best in Table~\ref{tab:SADresultskurdish}. Majority voting and logistic regression fusion multi-language results are called ASR\_Mul\_MV, and ASR\_Mul\_LR, respectively. For investigating the result of trainable ASR based and Pyannote models in in-domain scenario, the result of pre-trained Phn\_Rec and WebRTC models are not shown in Table~\ref{tab:SADresultskurdish}.

For in-domain experiment, temporal smoothing parameters are tuned using the in-domain development set. Here, w.r.t. ASR\_SingleBest model, ASR\_Mul\_LR improved the DetER for 1.2 \%. This LR fusion caused to decrease the miss detection with increasing the false alarm. ASR based SAD showed comparable performance w.r.t. the Pyannote model. Using different DNN architectures and temporal smoothing methods are the main reasons for observing the difference in the performance of these two systems. 
% \begin{figure}[h]
%     \centering
%     \includegraphics[width=0.45\textwidth, trim={0cm, 0cm, 0cm, 0cm}, ]{fig/det.pdf}
%     \caption{Detection error tradeoff (DET) curve of the test sets using multi-lingual ASR based SAD.}
%     \label{fig:det}
% \end{figure}
\vspace{-3mm}
\begin{table}[h]
\caption{Comparison of SAD results on in-domain BabelKurdish\_eval set. ASR\_SingleBest, ASR\_Mul\_LR, and ASR\_Mul\_MV are multi-lingual ASR based SAD systems when single best system, logistic regression based, or majority voting based fusion is considered, respectively.}
\centering
% \resizebox{0.48\textwidth}{!}{%
\begin{tabular}{@{}lccc@{}}
\toprule
SAD Model & DetER (\%) & FA (\%) & Miss (\%) \\
\midrule
ASR\_SingleBest & 19.9 & \bf{4.0} & 15.9 \\ 
ASR\_Mul\_LR & 18.7 & 5.2 & 13.5 \\ 
ASR\_Mul\_MV & 19.3 & 5.6 & 13.7 \\ 
% Phn\_Rec & 24.3 & 2.9 & 21.4 \\
% WebRTC & 31.7 & 4.1 & 27.6 \\ 
Pyannote & \bf{18.1} & 5.9 & \bf{12.2} \\ 
\bottomrule
\end{tabular}
%  }
\label{tab:SADresultskurdish}
\end{table}
%\vspace{-1mm}

% \begin{figure}[h]
%     \centering
%     \includegraphics[width=0.48\textwidth, trim={18mm, 0cm, 2cm, 0cm}, ]{fig/comp_plot.png}
%     \caption{Comparison of SAD segments of a sample audio file from evaluation part of Ester2 dataset. Pyannote segments show more FA. WebRTC segments show more FA and Miss. Phn\_Rec segments show more Miss.}
%     \label{fig:comp}
  
% \end{figure}

\vspace{-4mm}
\subsection{Out-of-domain evaluation}
\label{subsec:outofdomain}

Comparison of SAD results on out-of-domain LiveATC evaluation set is shown in the Table~\ref{tab:SADresultsatc}. Here ASR\_SingleBest and ASR\_Mul\_MV models are not using any in-domain data. 
ASR\_Mul\_LR model is trained using the in-domain development set. Without considering the ASR\_Mul\_LR model, ASR\_SingleBest and ASR\_Mul\_MV models, significantly outperformed the baseline models based on DetER performance measure. Training the multi-lingual AM model is one of the reasons for observing good result in the ASR\_SingleBest model. The ASR\_Mul\_LR model outperformed the ASR\_SingleBest model with relative improvement of 4.0\% on DetER performance measure. Comparison of SAD results on out-of-domain Ester2 evaluation set is shown in the Table~\ref{tab:SADresultsester}. In this out-of-domain set we observed the same pattern and based on DetER performance measure, the proposed model significantly outperformed the baselines. The ASR\_Mul\_LR model outperformed the ASR\_SingleBest model with relative improvement of 38.4\% on DetER performance measure. Based on the observed results, the proposed multi-lingual ASR based SAD, showed strong \emph{generalization ability}. We believe that training procedure as a multi-task learning system has the main effect on achieving this \emph{generalization ability}. In addition, having small in-domain dataset improves the performance of the proposed method.
% \vspace{-4mm}

\begin{table}[h]
\caption{Comparison of SAD results on out-of-domain LiveATC evaluation set. }
\centering
% \resizebox{0.48\textwidth}{!}{%
\begin{tabular}{@{}lccc@{}}
\toprule
SAD Model & DetER (\%) & FA (\%) & Miss (\%) \\
\midrule
ASR\_SingleBest & 10.1 & 4.9 & 5.2 \\ 
ASR\_Mul\_LR & \bf{9.7} & 6.1 & \bf{3.6} \\ 
ASR\_Mul\_MV & 11.1 & \bf{4.3} & 6.8 \\ 
Phn\_Rec & 20.1 & 4.6 & 15.5 \\
WebRTC & 16.5 & 9.4 & 7.1 \\ 
Pyannote & 13.8 & 10.1 & 3.7 \\ 
\bottomrule
\end{tabular}
%  }
\label{tab:SADresultsatc}
\end{table}
\vspace{-1mm}

\begin{table}[h]
\caption{Comparison of SAD results on out-of-domain Ester2 evaluation set.}
\centering
% \resizebox{0.48\textwidth}{!}{%
\begin{tabular}{@{}lccc@{}}
\toprule
SAD Model & DetER (\%) & FA (\%) & Miss (\%) \\
\midrule
ASR\_SingleBest & 5.2 & 4.7 & 0.5 \\ 
ASR\_Mul\_LR & \bf{3.2} & \bf{2.3} & 0.9 \\ 
ASR\_Mul\_MV & 4.7 & 4.2 & 0.5 \\ 
Phn\_Rec & 6.4 & 3.9 & 2.5 \\
WebRTC & 11.8 & 6.5 & 5.3 \\ 
Pyannote & 7.4 & 7.3 & \bf{0.1} \\ 
\bottomrule
\end{tabular}
%  }
\label{tab:SADresultsester}
\end{table}

% \vspace{-6mm}
\section{Conclusions}
\label{sec:conclusions}
Contextual information is important for training a robust SAD system, specially at noisy sets. In this paper, we trained the SAD system using multi-lingual ASR model. This ASR model was trained with LF-MMI loss on multi-task architecture which provides a much more scalable approach to develop AM. Decision for detecting speech/non-speech frames is based on index of maximum output posterior. Majority voting and logistic regression were applied to fuse the language-dependent decisions. We observed the significant improvement w.r.t. baselines on out-of-domain Ester2 and LiveATC evaluation sets. More specifically, for Ester2 dataset, the proposed SAD method outperformed the WebRTC, Phn\_Rec, and Pyannote BLSTM SAD models by absolute 7.1\%, 1.7\%, and 2.7\% in DetER respectively. Similarly, w.r.t. WebRTC, Phn\_Rec, and Pyannote BLSTM SAD models, respectively, we obtained an absolute improvement  6.4\%, 10.0\%, and 3.7\% in DetER on LiveATC dataset. In addition, using small development set in logistic regression method, further improved the performance of the proposed SAD system. In in-domain experiments, with tuning the temporal smoothing parameters we observed comparable result w.r.t. the Pyannote model. 
\vspace{-3mm}
\section{Acknowledgement}
\label{sec:ack}
This work was supported by the European Union’s Horizon 2020 research and innovation programme under grant agreement No. 833635 (ROXANNE: Real time network, text, and speaker analytics for combating organised crime).
The research is also partially based upon the work supported by the Office of
the Director of National Intelligence (ODNI), Intelligence Advanced Research
Projects Activity (IARPA), via AFRL Contract \#FA8650-17-C-9116. The views and
conclusions contained herein are those of the authors and should not be
interpreted as necessarily representing the official policies or endorsements,
either expressed or implied, of the ODNI, IARPA, or the U.S. Government. The
U.S. Government is authorized to reproduce and distribute reprints for
Governmental purposes notwithstanding any copyright annotation thereon.

\bibliographystyle{IEEEtran}

\bibliography{mybib}

\begin{thebibliography}{10}
\providecommand{\url}[1]{#1}
\csname url@samestyle\endcsname
\providecommand{\newblock}{\relax}
\providecommand{\bibinfo}[2]{#2}
\providecommand{\BIBentrySTDinterwordspacing}{\spaceskip=0pt\relax}
\providecommand{\BIBentryALTinterwordstretchfactor}{4}
\providecommand{\BIBentryALTinterwordspacing}{\spaceskip=\fontdimen2\font plus
\BIBentryALTinterwordstretchfactor\fontdimen3\font minus
  \fontdimen4\font\relax}
\providecommand{\BIBforeignlanguage}[2]{{%
\expandafter\ifx\csname l@#1\endcsname\relax
\typeout{** WARNING: IEEEtran.bst: No hyphenation pattern has been}%
\typeout{** loaded for the language `#1'. Using the pattern for}%
\typeout{** the default language instead.}%
\else
\language=\csname l@#1\endcsname
\fi
#2}}
\providecommand{\BIBdecl}{\relax}
\BIBdecl

\bibitem{sohn1999statistical}
J.~Sohn, N.~S. Kim, and W.~Sung, ``A statistical model-based voice activity
  detection,'' \emph{IEEE signal processing letters}, vol.~6, no.~1, pp. 1--3,
  1999.

\bibitem{sharma2019multi}
B.~Sharma, R.~K. Das, and H.~Li, ``Multi-level adaptive speech activity
  detector for speech in naturalistic environments.'' in \emph{INTERSPEECH},
  2019, pp. 2015--2019.

\bibitem{martinelli2020spiking}
F.~Martinelli, G.~Dellaferrera, P.~Mainar, and M.~Cernak, ``Spiking neural
  networks trained with backpropagation for low power neuromorphic
  implementation of voice activity detection,'' in \emph{ICASSP 2020-2020 IEEE
  International Conference on Acoustics, Speech and Signal Processing
  (ICASSP)}.\hskip 1em plus 0.5em minus 0.4em\relax IEEE, 2020, pp. 8544--8548.

\bibitem{dellaferrera2020bin}
G.~Dellaferrera, F.~Martinelli, and M.~Cernak, ``A bin encoding training of a
  spiking neural network based voice activity detection,'' in \emph{ICASSP
  2020-2020 IEEE International Conference on Acoustics, Speech and Signal
  Processing (ICASSP)}.\hskip 1em plus 0.5em minus 0.4em\relax IEEE, 2020, pp.
  3207--3211.

\bibitem{lee2020dual}
J.~Lee, Y.~Jung, and H.~Kim, ``Dual attention in time and frequency domain for
  voice activity detection,'' \emph{arXiv preprint arXiv:2003.12266}, 2020.

\bibitem{zheng2020mlnet}
Z.~Zheng, J.~Wang, N.~Cheng, J.~Luo, and J.~Xiao, ``Mlnet: An adaptive multiple
  receptive-field attention neural network for voice activity detection,''
  \emph{arXiv preprint arXiv:2008.05650}, 2020.

\bibitem{madikeri2020lattice}
S.~Madikeri, B.~Khonglah, S.~Tong, P.~Motlicek, H.~Bourlard, and D.~Povey,
  ``Lattice-free maximum mutual information training of multilingual speech
  recognition systems,'' \emph{Proc. of Interspeech 2020}, 2020.

\bibitem{karafiat2020but}
M.~Karafi{\'a}t, M.~K. Baskar, I.~Sz{\"o}ke, H.~K. Vydana, K.~Vesel{\`y},
  J.~{\v{C}}ernock{\`y} \emph{et~al.}, ``But opensat 2019 speech recognition
  system,'' \emph{arXiv preprint arXiv:2001.11360}, 2020.

\bibitem{madikeri2020pkwrap}
S.~Madikeri, S.~Tong, J.~Zuluaga-Gomez, A.~Vyas, P.~Motlicek, and H.~Bourlard,
  ``Pkwrap: a pytorch package for lf-mmi training of acoustic models,''
  \emph{arXiv preprint arXiv:2010.03466}, 2020.

\bibitem{kleinbaum2002logistic}
D.~G. Kleinbaum, K.~Dietz, M.~Gail, M.~Klein, and M.~Klein, \emph{Logistic
  regression}.\hskip 1em plus 0.5em minus 0.4em\relax Springer, 2002.

\bibitem{chuangsuwanich2011robust}
E.~Chuangsuwanich and J.~Glass, ``Robust voice activity detector for real world
  applications using harmonicity and modulation frequency,'' in \emph{Twelfth
  Annual Conference of the International Speech Communication Association},
  2011.

\bibitem{misra2012speech}
A.~Misra, ``Speech/nonspeech segmentation in web videos,'' in \emph{Thirteenth
  Annual Conference of the International Speech Communication Association},
  2012.

\bibitem{zazo2016feature}
R.~Zazo~Candil, T.~N. Sainath, G.~Simko, and C.~Parada, ``Feature learning with
  raw-waveform cldnns for voice activity detection,'' 2016.

\bibitem{ng2012developing}
T.~Ng, B.~Zhang, L.~Nguyen, S.~Matsoukas, X.~Zhou, N.~Mesgarani, K.~Vesel{\`y},
  and P.~Mat{\v{e}}jka, ``Developing a speech activity detection system for the
  darpa rats program,'' in \emph{Thirteenth annual conference of the
  international speech communication association}, 2012.

\bibitem{veisi2012hidden}
H.~Veisi and H.~Sameti, ``Hidden-markov-model-based voice activity detector
  with high speech detection rate for speech enhancement,'' \emph{IET signal
  processing}, vol.~6, no.~1, pp. 54--63, 2012.

\bibitem{enqing2002applying}
D.~Enqing, L.~Guizhong, Z.~Yatong, and Z.~Xiaodi, ``Applying support vector
  machines to voice activity detection,'' in \emph{6th International Conference
  on Signal Processing, 2002.}, vol.~2.\hskip 1em plus 0.5em minus 0.4em\relax
  IEEE, 2002, pp. 1124--1127.

\bibitem{zhang2012deep}
X.-L. Zhang and J.~Wu, ``Deep belief networks based voice activity detection,''
  \emph{IEEE Transactions on Audio, Speech, and Language Processing}, vol.~21,
  no.~4, pp. 697--710, 2012.

\bibitem{ivry2019voice}
A.~Ivry, B.~Berdugo, and I.~Cohen, ``Voice activity detection for transient
  noisy environment based on diffusion nets,'' \emph{IEEE Journal of Selected
  Topics in Signal Processing}, vol.~13, no.~2, pp. 254--264, 2019.

\bibitem{chang2018temporal}
S.-Y. Chang, B.~Li, G.~Simko, T.~N. Sainath, A.~Tripathi, A.~van~den Oord, and
  O.~Vinyals, ``Temporal modeling using dilated convolution and gating for
  voice-activity-detection,'' in \emph{2018 IEEE International Conference on
  Acoustics, Speech and Signal Processing (ICASSP)}.\hskip 1em plus 0.5em minus
  0.4em\relax IEEE, 2018, pp. 5549--5553.

\bibitem{hughes2013recurrent}
T.~Hughes and K.~Mierle, ``Recurrent neural networks for voice activity
  detection,'' in \emph{2013 IEEE International Conference on Acoustics, Speech
  and Signal Processing}.\hskip 1em plus 0.5em minus 0.4em\relax IEEE, 2013,
  pp. 7378--7382.

\bibitem{zhang2015boosting}
X.-L. Zhang and D.~Wang, ``Boosting contextual information for deep neural
  network based voice activity detection,'' \emph{IEEE/ACM Transactions on
  Audio, Speech, and Language Processing}, vol.~24, no.~2, pp. 252--264, 2015.

\bibitem{zhang2014boosted}
X.~L. Zhang and D.~Wang, ``Boosted deep neural networks and multi-resolution
  cochleagram features for voice activity detection,'' in \emph{Fifteenth
  annual conference of the international speech communication association},
  2014.

\bibitem{byers2019open}
F.~R. Byers, J.~G. Fiscus, S.~O. Sadjadi, G.~A. Sanders, and M.~A. Przybocki,
  ``Open speech analytic technologies pilot evaluation opensat pilot,'' Tech.
  Rep., 2019.

\bibitem{hadian2018end}
H.~Hadian, H.~Sameti, D.~Povey, and S.~Khudanpur, ``End-to-end speech
  recognition using lattice-free mmi.'' in \emph{Interspeech}, 2018, pp.
  12--16.

\bibitem{Bredin2020}
H.~{Bredin}, R.~{Yin}, J.~M. {Coria}, G.~{Gelly}, P.~{Korshunov},
  M.~{Lavechin}, D.~{Fustes}, H.~{Titeux}, W.~{Bouaziz}, and M.-P. {Gill},
  ``{pyannote.audio: neural building blocks for speaker diarization},'' in
  \emph{ICASSP 2020, IEEE International Conference on Acoustics, Speech, and
  Signal Processing}, Barcelona, Spain, May 2020.

\bibitem{schwarz2006hierarchical}
P.~Schwarz, P.~Matejka, and J.~Cernocky, ``Hierarchical structures of neural
  networks for phoneme recognition,'' in \emph{2006 IEEE International
  Conference on Acoustics Speech and Signal Processing Proceedings},
  vol.~1.\hskip 1em plus 0.5em minus 0.4em\relax IEEE, 2006, pp. I--I.

\bibitem{matejka2007but}
P.~Matejka, L.~Burget, O.~Glembek, P.~Schwarz, V.~Hubeika, M.~Fapso,
  T.~Mikolov, and O.~Plchot, ``But system description for nist lre 2007,'' in
  \emph{Proc. 2007 NIST Language Recognition Evaluation Workshop}, 2007, pp.
  1--5.

\end{thebibliography}

% \begin{thebibliography}{9}
% \bibitem[1]{Davis80-COP}
%   S.\ B.\ Davis and P.\ Mermelstein,
%   ``Comparison of parametric representation for monosyllabic word recognition in continuously spoken sentences,''
%   \textit{IEEE Transactions on Acoustics, Speech and Signal Processing}, vol.~28, no.~4, pp.~357--366, 1980.
% \bibitem[2]{Rabiner89-ATO}
%   L.\ R.\ Rabiner,
%   ``A tutorial on hidden Markov models and selected applications in speech recognition,''
%   \textit{Proceedings of the IEEE}, vol.~77, no.~2, pp.~257-286, 1989.
% \bibitem[3]{Hastie09-TEO}
%   T.\ Hastie, R.\ Tibshirani, and J.\ Friedman,
%   \textit{The Elements of Statistical Learning -- Data Mining, Inference, and Prediction}.
%   New York: Springer, 2009.
% \bibitem[4]{YourName17-XXX}
%   F.\ Lastname1, F.\ Lastname2, and F.\ Lastname3,
%   ``Title of your INTERSPEECH 2021 publication,''
%   in \textit{Interspeech 2021 -- 20\textsuperscript{th} Annual Conference of the International Speech Communication Association, September 15-19, Graz, Austria, Proceedings, Proceedings}, 2020, pp.~100--104.
% \end{thebibliography}

\end{document}